\documentclass[prl,aps,showpacs,twocolumn]{revtex4}
\usepackage{amssymb}

\usepackage{graphicx}

\begin{document}

\title{Twirling DNA Rings -\ Swimming Nanomotors Ready for a Kickstart}
\author{Igor M. Kuli\'{c}, Rochish Thaokar and Helmut Schiessel}
\affiliation{Max-Planck-Institut f\"{u}r Polymerforschung, Theory
Group, POBox 3148, D 55021 Mainz, Germany}

\date{\today}

\begin{abstract}

We propose a rotary DNA nanomachine that shows a continuous
rotation with a frequency of $10^2 -10^4$ Hz. This motor consists
of a DNA ring whose elastic features are tuned such that it can be
externally driven via a periodic temperature change. As a result
the ring propels itself through the fluid with a speed up to
microns per second.

\end{abstract}

\pacs{05.40.-a, 47.15.Gf, 87.15.He}

\maketitle

The long lasting dream of scaling mechanical devices and machines
down to the nanoscale (as popularized by R. Feynman
\cite{Feynman61} and carried on by several visionary groups
worldwide \cite{nanogroups}) continues to fire the imagination of
researchers -- now in the third generation. Among many
experimental difficulties that appear in this context, choosing
the proper material for the assembly of a nanodevice turns out to
be crucial. Important material requirements are: stability,
self-assembly ability, modularity, replicability, switchability,
experimental tractability. Presently one of the most promising
materials fulfilling those requirements is DNA \cite{Seeman04}.
Assemblies based on DNA hybridisation chemistry
\cite{Yurke00,Yan02,Turberfield03,Sherman04} as well as
conformational DNA transitions \cite{Mao99} were successfully
exploited to generate periodically switchable nanodevices.
However, despite their beauty and conceptual originality all of
these devices suffer one major problem: the large kinetic barriers
involved in the switching process boost their switching time per
cycle to $\sim 10^{3}$ s, four orders of magnitude slower than
their natural counterparts (biological molecular motors). A
natural question arises then: Can one achieve \textit{subsecond}
switching times with a DNA nanodevice? Can such a device be
operated in some manner to \textit{swim directionally and faster
than} $1\mu \mbox{m/s}$? In this letter we show theoretically the
principal feasibility of such DNA nanomachines.

\begin{figure}[tbp]
\includegraphics[width=0.8\linewidth]{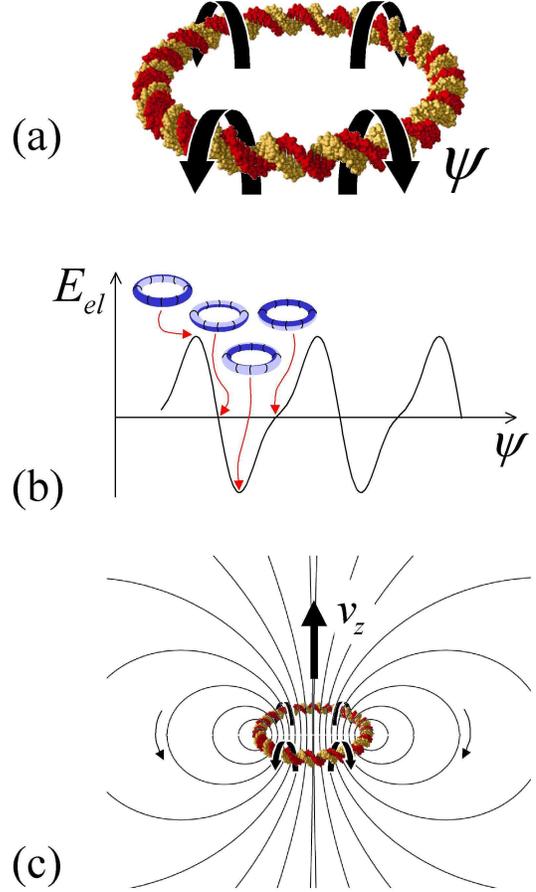}
\caption{The operation principle of the DNA-minicircle propeller:
(a) The twirling degree of freedom. (b) The elastic energy as a
ratchet potential. (c) The flowfield around the twirling ring
induces its translational velocity $v_z$.} \label{ring}
\end{figure}

Let us in the following propose a surprisingly simple nanomotor: a
DNA miniplasmid, cf. Fig.~1(a). We will show that despite its
structural simplicity a miniplasmid can be turned into a
nanomachine able to produce fN forces and self-propelling at
speeds of several microns per second through the fluid. In order
to run the plasmid as a motor we use here the Euler-angle $\psi $
(cf. Fig.~1(a)) as the relevant degree of freedom
\cite{footnote1}. The main idea now is to induce a directed
current $\langle \dot{\psi}\rangle $ -- in a manner similar to the
rotation of a closed rubber tube around its central circular axis
-- via non-equilibrium fluctuations and the ratchet effect
\cite{Juelicher97,Reimann02}, cf.~Fig.~1(b). As a result the
twirling ring generates a hydrodynamic flowfield (shown in
Fig.~1(c)) that remarkably induces an efficient self-propulsion of
the motor as detailed below.

The elastic distortion energy of a DNA\ ring with radius $R$
parametrized by the arc length parameter $s$ will in general be
described by three Euler angles $ \theta \left( s\right) ,\phi
\left( s\right) $ and $\psi \left( s\right) $ via
$E_{el}=\frac{1}{2}k_{B}T\int_{0}^{2\pi
R}\sum_{i=1,2,3}l_{i}\left( \omega _{i}-\kappa _{i}\right) ^{2}ds$
with $\omega _{1}=\phi ^{\prime }\sin \theta \sin \psi +\theta
^{\prime }\cos \psi $, $\omega _{2}=\phi ^{\prime }\sin \theta
\cos \psi -\theta ^{\prime }\sin \psi $ and $\omega _{3}=\phi
^{\prime }\cos \theta +\psi ^{\prime }$ \cite{Love}. Here $l_{1}$
and $l_{2}$ are the two principal bending persistence lengths and
$\kappa _{1}$ and $\kappa _{2}$ are intrinsic curvatures in two
corresponding perpendicular directions. $l_{3}$ denotes the twist
persistence length and -- for simplicity -- we choose $\kappa
_{3}=0$. The parameters $\kappa _{i}$ and $l_{i}$ reflect the
anisotropic bendability as well as intrinsic bendedness of the
plasmid; here for simplicity we assume them to be independent of
the arc-length throughout the molecule (e.g. for a 10 basepair
(bp) periodic sequence). For the case of DNA minicircles of short
length ($2\pi R\lesssim l_{i})$ and with constant $\kappa _{i}$
and $l_{i}$ fulfilling the weak bending anisotropy condition $\max
\left\{ \left\vert l_{1}-l_{2}\right\vert /R,l_{1}\kappa
_{1},l_{2}\kappa _{2}\right\} $ $\ll l_{3}/R$ only the
conformations close to the circular untwisted state will
contribute, i.e., those close to $\theta \left( s\right) =\pi /2,$
$\phi \left( s\right) =s/R$ and $\psi \left( s\right) =const$.
This leads then to the required ratchet potential acting on $
\psi$:
\begin{equation}
\frac{E_{el}\left( \psi \right) }{\pi k_{B}T}=
\frac{l_{1}-l_{2}}{2R} \cos \left( 2\psi \right) +2l_{1}\kappa
_{1}\cos \psi -2l_{2}\kappa _{2}\sin \psi \label{E_el}
\end{equation}
From Eq.~\ref{E_el} we see that for generating a ratchet potential
we need both nonzero bending anisotropy, $l_{1}-l_{2} \ne 0$, as
well as non-vanishing intrinsic curvatures, $\kappa _{1,2} \ne 0$.
The inset in Fig.~2 demonstrates that reasonable small values of
anisotropy and intrinsic curvature can induce a well-defined
ratchet potential.

The Fokker-Planck equation describing the time evolution of the
probability density $P\left( \psi ,t\right) $ of the Euler angle
$\psi $ writes
\begin{equation}
\zeta \frac{\partial P}{\partial t}=\frac{\partial }{\partial \psi
}\left( \frac{\partial E_{el}}{\partial \psi
}P+k_{B}T\frac{\partial P}{\partial \psi }\right)
\label{Langevin2}
\end{equation}
with the friction constant $\zeta $ that we will compute below. As
a source of non-equilibrium we will choose here a time-dependent
variation of temperature $T\left( t\right) $,
cf.~Ref.~\cite{Reimann96}.

\begin{figure}[tbp]
\includegraphics[width=0.9\linewidth]{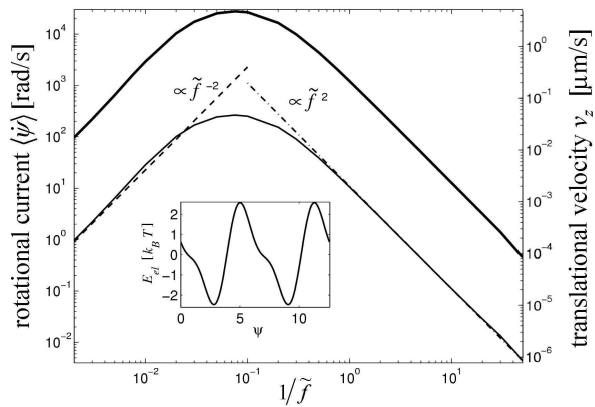}
\caption{The rotational current $\langle \dot{\psi}\rangle $ and
the induced translational velocity $v_z$ as a function of the
dimensionless frequency $\tilde{f}$ of the temperature (potential)
oscillations. The DNA ring has the following parameters: $R=10nm,$
$r_{0}=1nm$ (typical DNA minicircle), $l_{1}=45$ nm, $l_{2}=50$
nm, $\kappa _{1}=\kappa _{2}=(200\mbox{nm})^{-1}$ leading to the
ratchet potential displayed in the inset. Displayed are the
asymptotic expressions, Eqs.~\ref{Current} (dashed line) and
\ref{Current2} (dashed-dotted line) together with the numerical
solution of Eq.~\ref{Langevin2} (thin line) for a temperature
ratchet with $A_T=0.03$. The thick solid line corresponds to an
oscillating potential ratchet with $A_E =0.3$. See text for
details.} \label{fig2}
\end{figure}

Before we compute the friction constant $\zeta $ we need to shed
some light on the low Reynolds number hydrodynamics of the
twirling DNA ring. The latter turns out to be peculiarly related
to the inviscid (ideal) fluid vortices (rings of smoke) and as a
matter of fact both of them propagate in analogous manner. To see
this we first remark that for a reasonable ring radius $R=10$ nm
(a typical miniplasmid of $\approx 200$ bp) and the DNA helix
radius $r_{0}=1$ nm the slender body approximation
\cite{Johnson79} is valid with the slenderness paramenter
$\varepsilon =r_{0}/R=0.1$. In the spirit of the slender body
theory one approximates the flow-field around the twirling ring by
superimposing rotlets \cite{Chwang90} $\mathbf{u}_{rot}\left(
\mathbf{x};s\right) =\Gamma \frac{d \mathbf{c}\left( s\right)
}{ds}\times \left( \mathbf{x}-\mathbf{c}\left( s\right) \right)
/\left| \mathbf{c}\left( s\right) -\mathbf{x}\right| ^{3}$ placed
along the ring centerline $\mathbf{c}\left( s\right) $ with
arclength parameter $s$. The rotlet strength $\Gamma
=\frac{1}{2}\omega _{\mathbf{c} }r_{0}^{2}$ is given in terms of
the angular velocity $\omega _{\mathbf{c}}$ of the ring about
$\mathbf{c}\left( s\right) $. The full velocity profile is then
given by $\mathbf{u}\left( \mathbf{x}\right) =\int_{0}^{2\pi
R}\mathbf{u }_{rot}\left( \mathbf{x};s\right) ds$; cf. also the
stream lines around the rotating ring shown in Fig.~1(c). When
integrating $\mathbf{u}\left( \mathbf{x}\right) $ over the DNA
ring (slender torus) surface in the limit of small $r_{0}/R$ one
obtains a net translational velocity in the z-direction:
\begin{equation}
v_{z}\left( \omega _{\mathbf{c}}\right) =\frac{r_{0}^{2}}{2R}\left( \ln
\left( 8\frac{R}{r_{0}}\right) -\frac{1}{2}\right) \omega _{\mathbf{c}}
\label{v_Ring}
\end{equation}
The fact that Eq.~\ref{v_Ring} coincides with the well known
expression from ideal flow vortex theory \cite{Lamb32} should not
surprise if we recall that a rotlet $\mathbf{u}_{rot}\left(
\mathbf{x};s\right) $ is nothing else but the expression for the
velocity field of an ideal point vortex. But despite this
kinematic analogy between the twirling DNA and an ideal vortex
ring, dynamically they are quite different. The propagation of an
ideal vortex ring does not require any external forces/torques and
is governed by conservation of kinetic energy and momentum. In
sharp contrast to that the low Reynolds-number (Stokes) flow is
governed by dissipation and the motion of twirling DNA ring
requires the action of a torque $N_{\mathbf{c}}=8\pi
^{2}x_{0}^{2}\eta R\omega _{\mathbf{c}}$ ($\eta =10^{-3}$~Pa~s,
the water viscosity) about the central axis $\mathbf{c}$. The
latter expression can be verified by integrating the tangential
stresses generated by $\mathbf{u} \left( \mathbf{x}\right) $ over
the ring surface. More generally by virtue of the linearity of the
Stokes equations we can derive a resistance matrix $ \left(
M_{kl}\right) $ relating the angular velocity $\omega
_{\mathbf{c}}$ (about the circular axis $\mathbf{c}$) and velocity
$v_{z}$ (in the z-direction) with the corresponding external
torque $N_{\mathbf{c}}$ and force $F_{z}$ :
\begin{equation}
\left(
\begin{array}{c}
F_{z} \\
N_{\mathbf{c}}
\end{array}
\right) =4\pi ^{2}\eta \left(
\begin{array}{cc}
M_{11} & M_{12} \\
M_{21} & M_{22}
\end{array}
\right) \left(
\begin{array}{c}
v_{z} \\
\omega _{\mathbf{c}}
\end{array}
\right)  \label{Mobility}
\end{equation}
Combining the previous expressions obtained for $v_{z}\left(
\omega _{ \mathbf{c}}\right) $ and $N_{\mathbf{c}}\left( \omega
_{\mathbf{c}}\right) $ ($F_{z}=0$) together with the result of
Johnson and Wu \cite{Johnson79} for the drag on a \textit{rigid}
slender torus we obtain entries in the leading order: $ M_{11}=$
$2R\left( \ln 8/\varepsilon +1/2\right) ^{-1}$,
$M_{22}=2r_{0}^{2}R$ and $M_{12}=M_{21}=r_{0}^{2}\left( \ln
8/\varepsilon -1/2\right) \left( \ln 8/\varepsilon +1/2\right)
^{-1}$. Note that the symmetry of the resistance matrix being a
general feature of swimmers in the Stokes flow \cite{Purcell97}
provides a good check for the consistency of the involved
calculations. Having the mobility relation \ref{Mobility} one can
consider different types of motion, e.g.: (\textit{i}) The ring is
twirling freely ($N_{\mathbf{c}}=0$) and moving under the force $
F_{z}$ (or at fixed $v_{z}$). (\textit{ii}) The ring is prevented
from rotation ($ \omega _{\mathbf{c}}=0$) and moved by the force
$F_{z}$ (or at fixed $v_{z}$). (\textit{iii}) The ring is held in
position ($v_{z}=0$) by imposing a force $F_{z}$ counterbalancing
the action of torque $N_{\mathbf{c}}$. (\textit{iv}) The DNA ring
is free to move (no external force applied, $F_{z}=0$) under an
externally imposed torque $N_{\mathbf{c}}$ (or at given $\omega
_{\mathbf{c}}$). Comparing, for instance, the velocities
$v_{z}^{(i)}$ and $v_{z}^{(ii)}$ from cases (\textit{i}) and
(\textit{ii}) one sees that
$v_{z}^{(i)}/v_{z}^{(ii)}-1=M_{12}^{2}/\det M_{kl}\sim \left(
\varepsilon /2\right) ^{2}\ln \left( 8/\varepsilon \right) $;
i.e., a ring with an isotropic DNA sequence (able of twirling)
settles faster than a ring with very high barriers to twirling
($\omega _{\mathbf{c} }=0$), though in practice the difference is
negligible, for instance $ \sim 1\%$ for $\varepsilon =0.1$. By
comparing cases (\textit{iii}) and (\textit{iv}) we conclude that
a ring twirling at fixed $\omega _{\mathbf{c}}$ and forced not to
translate requires a slightly larger torque
$N_{\mathbf{c}}^{iii}\left( \omega _{\mathbf{c} }\right) $ than
the unconstrained freely translating ring with $N_{\mathbf{c}
}^{iv}\left( \omega _{\mathbf{c}}\right) $. However the relative
difference is again small and of order $O\left( \varepsilon
^{2}\ln \left( 8/\varepsilon \right) \right) $. Therefore by
dropping this marginal correction to the leading order we obtain
in both cases (\textit{iii}) and (\textit{iv}) the angular
friction constant $\zeta =N_{\mathbf{c}}\left( \omega _{\mathbf{c}
}\right) /\omega _{\mathbf{c}}\approx 8\pi ^{2}\eta r_{0}^{2}R$,
the quantity that appeared above in Eq.~\ref{Langevin2}. Note that
the latter is the same (in our $\varepsilon \ll 1$ leading order
expansion) as for a straight cylinder with radius $r_{0}$ and
length $2\pi R$. Finally another interesting feature that can be
read off Eq.~\ref{Mobility} is the efficiency of the twirling ring
propulsion. The latter is independent of the mechanism of twirling
and can be defined as the ratio of the power $P_{0}=2\pi ^{2}\eta
M_{11}v_{z}^{2}$ dissipated by a (for simplicity) rigid ring
directly moved by a force as compared to the power
$P_{twirl}=\frac{1}{2}N_{\mathbf{c} }\omega _{\mathbf{c}}$
dissipated by twirling propulsion at the same translational speed.
For a ring with $R=10$ nm we have $P_{0}/P_{twirl} \approx 0.8\%$
which is comparable to the efficiency of bacterial propulsion by a
rotating flagellum \cite{Purcell97}.

Having determined the friction constant $\zeta $ we return to the
ratchet dynamics given by Eq.~\ref{Langevin2} with the twirling
potential Eq.~\ref {E_el}. To obtain the directed twirling
frequency $\omega _{\mathbf{c} }:=\langle
\dot{\psi}\rangle=-\frac{1}{\zeta} \langle \frac{\partial
E_{el}}{\partial \psi}P+k_B T \frac{\partial P}{\partial
\psi}\rangle$ we follow Ref.~\cite{Reimann96} by choosing a
periodic time dependent temperature variation as follows: $
T\left( t\right) =T_{0}\left[ 1+A_{T}\sin (2\pi f_{T}t)\right] $
with $T_{0}$ the mean temperature, $A_{T}$ the relative amplitude
and $f_{T}$ the frequency of the temperature oscillation. For the
case of $f_{T}$ sufficiently larger than the inverse of the
characteristic relaxation time $\tau_0=4\pi ^{2}\zeta/\left(
k_{B}T_{0} \right) $ of the twirling degree of freedom (but still
much smaller than the frequency of average thermal molecular
kicks) an $1/f_{T}$ asymptotic expansion for the current $\langle
\dot{\psi}\rangle $ can be employed \cite{footnote3}. After a long
calculation we obtain (for $f_{T}>f_{res}$) $\langle
\dot{\psi}\rangle$ up to terms of order $O\left( f
_{T}^{-3}\right)$:
\begin{equation}
\left\langle \dot{\psi}\right\rangle =\allowbreak \frac{12\pi
^{3}A_{T}^{2}\left( k_{B}T_{0}\right) ^{3}l_{1}l_{2}\kappa
_{1}\kappa _{2}\left( l_{2}-l_{1}\right) /R}{f_{T}^{2}\zeta
^{3}\int_{0}^{2\pi }d\psi
e^{-\frac{E_{el}\left( \psi \right) }{k_{B}T}}\int_{0}^{2\pi }d\psi e^{\frac{%
E_{el}\left( \psi \right) }{k_{B}T}}} \label{Current}
\end{equation}
From Eq.~\ref{Current} we see that for an isotropically bendable
DNA sequence $(l_{2}=l_{1})$ the directed current vanishes. The
same is true if the intrinsic curvature direction coincides with
one of the principal axes (i.e. if $\kappa _{1}$ or $\kappa _{2}$
vanish). Both observations are intuitive as in either case the
ratchet potential, Eq.~\ref{E_el}, becomes left-right symmetric
and the ratchet effect disappears.

The low frequency adiabatic limit is obtained from the asymptotic
expansion of $P( \tilde{\psi} ,\tilde{t}) $
($\tilde{\psi}=\psi/2\pi$, $\tilde{t}=t/\tau_0$) for small
parameter $\tilde{f} =f_{T}\tau _{0}$, i.e., $P\approx
P_{0}+\tilde{f}P_{1}+\tilde{f} ^{2}P_{2}$. Rather involved
calculations lead to \cite{footnote4}
\begin{equation}
\left\langle \dot{\psi}\right\rangle =-\frac{\tilde{f}^{2}}{ \tau
_{0}}\int_{0}^{1}d\tilde{t} \frac{1}{\overline{F}}\overline{F
\overrightarrow{\partial_{\tilde t}P_{1}}} \label{Current2}
\end{equation}
with $E( \tilde{\psi},\tilde{t}) =F( \tilde{\psi},\tilde{t})
^{-1}=e^{-E_{el}( \tilde{\psi}) /k_{B}T( \tilde{t}) }$ and the
abbreviations $\overrightarrow{\left( ...\right) }$ and
$\overline{\left( ...\right) }$ defined as in \cite{footnote3} but
with the integrations with respect to $\tilde \psi$. Furthermore
the density distributions $P_{0}$ and $P_{1}$ from the upper
expansion are given by $P_{0}=E/\overline{E}$ (Boltzmann
distribution in the adiabatic limit) and
$P_{1}=\frac{T_{0}}{T}E\left(
\overrightarrow{Fc_{1}}-\frac{1}{\overline{E}}\overline{E\overrightarrow{
Fc_{1}}}\right) $ with $c_{1}=\overrightarrow{\partial_{\tilde
t}{P}_{0}}-\frac{1}{
\overline{F}}\overline{F\overrightarrow{\partial_{\tilde
t}{P}_{0}}}$.

Equations \ref{Current} and \ref{Current2} together with
Eq.~\ref{E_el} and $\zeta =8\pi ^{2}\eta r_{0}^{2}R$ allows us to
get the twirling speed $\omega _{\mathbf{c }}=\langle \dot{
\psi}\rangle $ and by virtue of Eq.~\ref{v_Ring} the induced
translational velocity $v_{z}\left( \omega _{\mathbf{c}}\right) $
for arbitrary DNA elastic parameters $l_{i=1,2}$ and $\kappa
_{i=1,2}$.

How fast can we operate the twirling ring machine? We shall assume
some realistic parameter values for the DNA ring: $R=10$ nm,
$r_{0}=1$ nm (typical DNA minicircle) leading to $\zeta =2\cdot
10^{-7}k_{B}T$s. Furthermore we set $l_{1}=45$ nm, $l_{2}=50$ nm,
$\kappa _{1}=\kappa _{2}=(200\mbox{nm})^{-1}$ which corresponds to
a rather modest anisotropy and intrinsic curvature. For the
temperature variation amplitude we choose $\Delta T=\pm 10$ K,
i.e., $ A_{T}\approx 1/30$ (at room temperature $T_{0}=300$ K).
Figure 2 provides a log-log plot of the rotational current and the
corresponding drift speed of the ring as a function of the
dimensionless frequency $\tilde f$ of the temperature variation.
The thin solid curve gives the numerical result obtained from
Eq.~\ref{Langevin2}, the two straight lines correspond to the
analytical results for the two asymptotic cases,
Eqs.~\ref{Current} and \ref{Current2}. As can be seen from this
plot the two limits show a $\tilde{f}^{-2}$ and $\tilde{f}^2$
dependence, respectively, in accordance with Eqs.~\ref{Current}
and \ref{Current2}. The maximal rotational current is achieved in
the crossover region, namely $\omega _{\mathbf{c}} \approx 200
\mbox{rad}/\mbox{s}$ for $\tilde f \approx 10^{-1}$. Following
Eq.~\ref{v_Ring} this implies a translational velocity of
$v_{z}=50 \mbox{nm}/\mbox{s}$.

Such fast temperature oscillations are technically feasible and
might be most conveniently generated by adiabatic pressure
variations e.g. by ultrasound. Another promising method is to use
the inductive heating of metal nanocrystals that are covalently
attached to the DNA ring. In fact, this method has been
successfully used to control the hybridization behavior of DNA
\cite{Hamad02}. This might also point towards an alternative way
of driving the ratchet, namely via a periodic variation of the
elastic properties of the ring. Operating the system close to the
DNA duplex melting temperature is likely to induce strong
oscillations in the overall ring stiffness. The thick solid line
in Fig.~2 shows the rotational current obtained when the elastic
energy is varied as $\tilde
E_{el}\left(\psi,t\right)=E_{el}\left(\psi\right) \left(1+A_E
\sin\left(2\pi f_E t\right) \right)$ where we chose the relative
amplitude $A_E =0.3$. As can be seen from Fig.~2 the maximal
current of this oscillating potential ratchet occurs roughly at
the same frequency as that of the thermal ratchet but the value of
$\omega _{\mathbf{c}}$ is much higher, namely on the order of $2
\times 10^4 \mbox{rad}/\mbox{s}$ which implies a quite notable
translational velocity of $v_{z}=$\ $5 \mu \mbox{m}/\mbox{s}$. As
a comparison a typical bacterium moves at $30\mu
\mbox{m}/\mbox{s}$. Our ring ratchet (with oscillating potential)
resembles in many respects ''real'' biological nanomotors. Besides
its nanoscopic size (radius 10 nm), swimming efficiency (0.8\%)
and speed ($4\mu \mbox{m}/\mbox{s}$) it can generate forces and
torques close to that of biomolecular motors. Although the net
translational force resulting from Eq.~\ref{Mobility} $F_{z}=4\pi
^{2}\eta M_{12}\omega _{\mathbf{c}}\approx 0.6$ fN is comparably
small (due to cancelling of most of the stresses), the local
torque $N_{\mathbf{c}}=8\pi ^{2}\eta r_{0}^{2}R\omega
_{\mathbf{c}}\approx 0.004k_{B}T$ and the force acting at the DNA
surface $F_{loc}=N_{\mathbf{c}}/r_{0}=\zeta \omega _{\mathbf{c}
}/r_{0}\approx 16$ fN are significant if we consider the
simplicity of the mechanism behind.

From an experimental point of view one should be aware of the fact
that a ring (twirling or non-twirling) looses its initial
orientation almost instantaneously due to rotational diffusion.
The typical relaxation time scale of this process is on the order
$\eta R^3/\left( k_B T\right)$ (up to logarithmic corrections
\cite{Paul69}) which for a ring with $R=10$ nm leads to $10^{-7}$
s. That means that a single twirling ring in solution will not
perform any noticeable translational drift. A possible solution to
the problem is to put the ring on a "track", e.g. to thread it on
a straight DNA chain. Another promising direction is to prepare a
semi-dilute solution of such rings and then study their response
due to an induced twirling. The resulting hydrodynamic flow can
lead to an attraction between the rings that might facilitate
interesting collective behavior of the nanomotors like the
formation of columns of twirling rings that drive a solvent flow
through the resulting channel.

Acknowledgements: The authors thank Markus Deserno, Vladimir
Lobaskin and Pim Schravendijk for discussions.


\begin{thebibliography}{99}

\bibitem{Feynman61} R. P. Feynman in, \textit{Miniaturization},
edited by H. D. Gilbert (Reinhold, New York, 1961), pp. 282-296.

\bibitem{nanogroups} R. A. Bissell, E. Cordova, A. E. Kaifer, and
J. F. Stoddart, Nature (London) \textbf{369}, 133 (1994); T. R.
Kelly, H. D. Silva and R. A. Silva, Nature (London) \textbf{401},
150 (1999); N. Koumara, R. W. J. Zijlstra, R. A. van Delden, N.
Harada, and B. L. Feringa, Nature (London) \textbf{401}, 152
(1999).

\bibitem{Seeman04}  N. C. Seeman, Sci. American \textbf{290}, 64 (2004).

\bibitem{Yurke00}  B. Yurke, A. J. Turberfield, A. P. Mills,
F. C. Simmel and J. L. Neumann,  Nature \textbf{406}, 605 (2000).

\bibitem{Yan02}  H. Yan, X. Zhang, Z. Shen, and N. C. Seeman,
Nature \textbf{415}, 62 (2002).

\bibitem{Turberfield03}  A. J. Turberfield, J. C. Mitchell, B. Yurke,
A. P. Mills, M. I. Blakey, and F. C. Simmel,  Phys. Rev. Lett.
\textbf{90}, 118102 (2003).

\bibitem{Sherman04} W. B. Sherman and N. C. Seeman, Nano Lett.
\textbf{4}, 1203 (2004).

\bibitem{Mao99}  C. Mao, W. Sun, Z. Shen, and N. C. Seeman,
Nature \textbf{397}, 144 (1999).

\bibitem{footnote1} It is easy to show that provided that the plasmid
length is $\lesssim $ $ l_{P}$ (persistence length) all other
degrees of freedom besides the $\psi $ motion can be neglected.

\bibitem{Juelicher97}A. Ajdari and J. Prost, C. R. Acad. Sci. (Paris) \textbf{315},
1635 (1992); M. O. Magnasco, Phys. Rev. Lett. \textbf{71}, 1477
(1993) ; F. J\"ulicher, A. Ajdari, and J. Prost, Rev. Mod. Phys.
\textbf{69}, 1269 (1997).

\bibitem{Reimann02} P. Reimann, Phys. Rep. \textbf{361}, 57
(2002).

\bibitem{Love} A. E. Love, \textit{A Treatise on the Mathematical
Theory of Elasticity} 4th edition, (Dover, New York, 1944).

\bibitem{Reimann96} P. Reimann, R. Bartussek, R. H\"aussler, and
P. H\"anggi, Phys. Lett. A \textbf{215}, 26 (1996).

\bibitem{Johnson79} R. E. Johnson, and T. Y. Wu, J. Fluid Mech.
\textbf{95}, 263 (1979).

\bibitem{Chwang90} A. T. Chwang, and W.-S. Hwang, Phys. Fluids A
\textbf{2}, 1309 (1990).

\bibitem{Lamb32} H. Lamb, \textit{Hydrodynamics} 6th edition, (Cambridge Univ.
Press, 1993).

\bibitem{Purcell97} E. M. Purcell, Proc. Natl. Acad. Sci. USA
\textbf{94}, 11307 (1997).

\bibitem{footnote3}
Note that the explansion given in Ref.~\cite{Reimann02,Reimann96}
is not invariant with respect to time shifts. The correct time
dependence term in Eq. (2.58) of Ref.~\cite{Reimann02} should read
$\overline{\overrightarrow{ \Delta }^{2}}-\left(
\overline{\overrightarrow{\Delta }}\right) ^{2}$ instead of
$\overline{\overrightarrow{\Delta }^{2}}$ with $\Delta \left(
h\right) =1-\frac{T\left( 2\pi h\right) }{T_{0}}$. Here we
introduce the abbreviations $\overrightarrow{\left( ...\right)
}=\int_{0}^{x}\left( ...\right) dx^{\prime }$ and
$\overline{\left( ...\right) } =\int_{0}^{1}\left( ...\right)
dx^{\prime }$ for definite and indefinite integration.

\bibitem{footnote4} Note that the $O(\tilde{f})$-term
provided in Ref.~\cite{Reimann96}, Eq.~(18), \textit{always}
vanishes identically (integrated surface term) so that the second
order term, Eq.~\ref{Current2}, constitutes the first
non-vanishing contribution to the current.

\bibitem{Hamad02} K. Hamad-Schifferli, J. J. Schwartz, A. T.
Santos, S. Zhang, and J. M. Jacobson, Nature (London)
\textbf{415}, 152 (2002).

\bibitem{Paul69} E. Paul and R. M. Mazo, J. Chem. Phys.
\textbf{51}, 1102 (1969).

\end{thebibliography}
\end{document}